\documentclass[aip,graphicx]{revtex4-1}
\usepackage{epsfig}

\begin{document}


\title{Impressive Transport Properties of Be$_2$C Monolayer} 

\author{Gautam Sharma}
\email{gautam.iiser@gmail.com}

\affiliation{Department of Physics, Indian Institute of  Science Education and Research Pune, Pune, India}
\author{K.C. Bhamu}
\affiliation{PMC Division, CSIR-National Chemical Laboratory, Pune-411008, India}

\begin{abstract}
We present thermoelectric properties of Be$_2$C monolayer based on density functional theory and semi-classical Boltzmann transport theory. Electronic structure calculations predict this material as a semiconductor with a direct bandgap of 2.0 eV computed using Gaussian-attenuating Perdew-Burke-Ernzerhof (Gau-PBE) hybrid functional. 
The Gau-PBE band structure is used to compute transport properties by solving the Boltzmann transport equation under the constant relaxation time approximation. In this work, we have explicitly determined the relaxation time by studying the electron-phonon interactions in the system to estimate absolute transport coefficients. Our results show that the monolayer possesses a high power factor ($\sim$ 3.44 mW/mK$^2$), similar to the commercial TE materials doped-Bi$_2$Te$_3$ and PbTe, suggesting that Be$_2$C monolayer is a promising thermoelectric material.
\end{abstract}


\maketitle 

\section{INTRODUCTION}
Increasing consumption of natural energy resources demanded the scientific community to switch their
attention to sustainable energy resources. Waste heat management technology is one of the efficient and economical alternatives for power generation. Thermoelectric (TE) materials have gained tremendous attention for their ability to convert the waste heat ejected from automobiles and industries into electricity. The figure of merit, $ZT$ = $\alpha^2\sigma$T/$\kappa$, is the crucial quantity that describes a TE material's efficiency, where T, $\alpha$, $\sigma$, and $\kappa$ denote temperature, Seebeck coefficient, electrical conductivity, and total thermal conductivity, respectively. The total thermal conductivity is the sum of the electronic ($\kappa_e$) and lattice thermal conductivity ($\kappa_l$). $ZT$ can be maximized by increasing the power factor (PF, PF=$\alpha^2\sigma$) and simultaneously decreasing the thermal conductivity ($\kappa$). However, it is very challenging to achieve high PF because of the interdependence of the quantities involved. 
For example, if one tries to enhance $\sigma$ through doping the material, it reduces $\alpha$. Moreover, an increase in $\sigma$ also increases $\kappa_e$ as Wiedemann-Franz Law connects them. Therefore it is not easy to control these parameters independently and thus to optimize $ZT$ of a material.

Despite these challenges, TE performance of the material can be tuned by various ways like convergence of multiple bands,\cite{george}\cite{pbte_seebeck1} combination of light-heavy bands\cite{light-heavy_masses}, multivalley carrier pocket\cite{carrier_pocket} and reduced dimensionality.\cite{dj_low_dim} Dresselhaus \textit{et al.} have shown theoretically that low-dimensional systems would attain higher ZT than their bulk counterpart\cite{hicks_2d}\cite{hicks_1d}. Recent advances to improve ZT are centered around reducing dimensionality, for instance, monolayer Bi$_2$Te$_3$ have shown high ZT than its bulk counterpart\cite{bi2te3_th}\cite{bi2te3_exp}\cite{phosphorene_nano}.
Similarly, nanostructured materials have also shown better TE performance\cite{nano-structure_review}.

In this work, we study thermoelectric properties of Be$_2$C monolayer (Be$_2$C-ML) in its global minimum energy structure using density functional theory (DFT) based calculations and semi-classical Boltzmann transport theory (BTT). Be$_2$C-ML is found to be a semiconductor with a bandgap of 2.34 eV computed with Heyd-Scuseria-Ernzerhof (HSE) functional\cite{angwe_be2c}. In a computational study, Naseri \textit{et al.} have investigated electronic and optical properties of the Be$_2$C-ML under stress and strain\cite{be2c_optical}. Yeoh \textit{et al.} have studied Be$_2$C-ML as the anode material for Lithium-ion battery applications\cite{be2c_battery}.  Notably, Be and C elements are earth-abundant relative to elements (like Pb, Bi, and Te) used for commercial TE materials. Moreover, a molecular dynamics-based study has shown that the Be$_2$C-ML has a high melting point around 1500 K, and therefore, it has the potential to show stunning TE properties at high temperature\cite{angwe_be2c}. 

The electronic band structure shows that it has doubly degenerate valence bands with different band curvatures around the Brillouin zone center. Such a band feature is often desirable in TE materials as it provides both light and heavy hole pockets to enhance the transport properties\cite{angwe_be2c,light-heavy_masses}. To the best of our knowledge, TE properties of the Be$_2$C-ML are still waiting to uncover. These factors have motivated us to investigate the TE properties of the Be$_2$C-ML, which includes studying the electron-phonon interactions (EPIs) and thereby calculating the relaxation time of charge carriers in the system. The rest of the manuscript is organized as follows. Section \ref{cd} provides the details of computational methods used to investigate various properties. Section \ref{rd} discusses electronic structure, EPIs and TE properties of the Be$_2$C-ML. Conclusive remarks are presented in Section \ref{concl}.

\section{COMPUTATIONAL DETAILS}
\label{cd}
\subsection{DFT calculations}
First-principles calculations are performed with the Quantum ESPRESSO software, which is a plane-wave based implementation of density functional theory (DFT)\cite{QE-2009,QE-2017}. The electronic exchange-correlation potential is described by the generalized gradient approximation (GGA) as parametrized by Perdew, Burke, and Ernzerhof (PBE)\cite{pbe}. We have used the norm-conserving Trouiller-Martins pseudopotentials to describe the electron-ion interactions\cite{ncpp}. These pseudopotentials have been generated using 2s$^2$ 2p$^0$ 3d$^0$ 4f$^0$ and 2s$^2$ 2p$^2$ 3d$^0$ 4f$^0$ valence configurations for Be and C, respectively. The wave functions are expanded in a plane wave basis, defined by the kinetic energy cut-off of 85 Ry. Brillouin zone (BZ) integrations are performed using a Monkhorst-Pack $k$-point grid of (11$\times$11$\times$1) for the monolayer\cite{kpts}. We have included semi-empirical Grimme-D2 van der Waals (vdW) corrections in all the calculations\cite{vdw}. For a better estimate of the bandgap, we have performed hybrid calculations using the singularity free Gaussian-attenuating Perdew-Burke-Ernzerhof (Gau-PBE) functional\cite{gau}. The matrix elements of the Fock operator are evaluated with a (5$\times$5$\times$1) $q$-grid for the ML. We have used a vacuum of length L, 15 {\ AA}, to eliminate the spurious interactions due to periodic images present along the z-axis.
\subsection{Relaxation time ($\tau$) calculations:}
Within Migdal approximation, the electron self-energy ($\Sigma_{n,k}$=$\Sigma_{n,k}^{\prime}$+$\imath$ $\Sigma_{n,k}^{\prime\prime}$) is given by\cite{migdal_approx1}\cite{migdal_approx2}

 \begin{equation}
   \sum_{nk}  = \sum_{\mathrm{q}\nu,m} w_\mathrm{q} |g_{\mathrm{q}\nu,m}(k,\mathrm{q})|^2 
   \bigg [ \frac{n_{q\nu}+1-f_{mk+q}}{\epsilon_{nk}-\epsilon_{mk+q}-\hbar\omega_{q\nu}-i\delta} +
   \frac{n_{q\nu}+f_{mk+q}}{\epsilon_{nk}-\epsilon_{mk+q}+\hbar\omega_{q\nu}-i\delta} \bigg]
   \label{epw-eq}
   \end{equation}

where $g_{q\nu,m}$ is electron-phonon (el-ph) matrix element. $n_{q\nu}$ and $f_{mk}$ are the Bose and Fermi occupation factors respectively. $\omega_{q\nu}$ is the phonon frequency for mode $\nu$ and wavevector q and $\epsilon_{nk}$ are the Kohn-Sham eigenvalues for band n and wavevector k. The imaginary part of self-energy of the electron (Im$\Sigma_{n,k}$) is computed with the EPW package{\cite{epw}}.
The Im$\Sigma_{n,k}$ is related to the relaxation time as follows:
\begin{equation}
 \tau_{n,k} = \frac{\hbar}{2\textnormal{Im}\Sigma_{n,k}}    \label{tau_sigma}
\end{equation}
where $\hbar$ is the reduced Planck's constant.
We calculate the electron self-energy using PBE functional, firstly, we compute el-ph matrix elements (g$_{mn}^{\nu}$) on a coarse ($18\times18\times1$) $k$-grid and (9$\times$9$\times$1) $q$-grid. To obtain converged results, the coarse $k$ and $q$-grids are interpolated on a (25$\times$25$\times$1) $k$-grid and a dense ($280\times280\times1$) $q$-grid respectively.

\subsection{Boltzmann transport calculations:}

Transport coefficients such as Seebeck coefficient ($\alpha$), electrical conductivity ($\sigma$) can be calculated using semi-classical Boltzmann Transport equation (BTE)\cite{mahan}. BoltzTraP package is used to solve the BTE to
provide TE coefficients within the rigid band approximation (RBA) and constant relaxation time approximations (CRTA)\cite{bt}. Within RBA, doping and temperature effects on the band energies are ignored. The chemical potential ($\mu$) is shifted above/below for electron/hole doping. We have found the convergence in transport coefficients computed based on the electronic structure calculations performed with Gau-PBE functional using a dense (35$\times$35$\times$1) $k$-grid.

\section{RESULTS AND DISCUSSIONS}
\label{rd}
\subsection{Crystal structure}
\begin{figure}[h]
\centering
\includegraphics[scale=0.5]{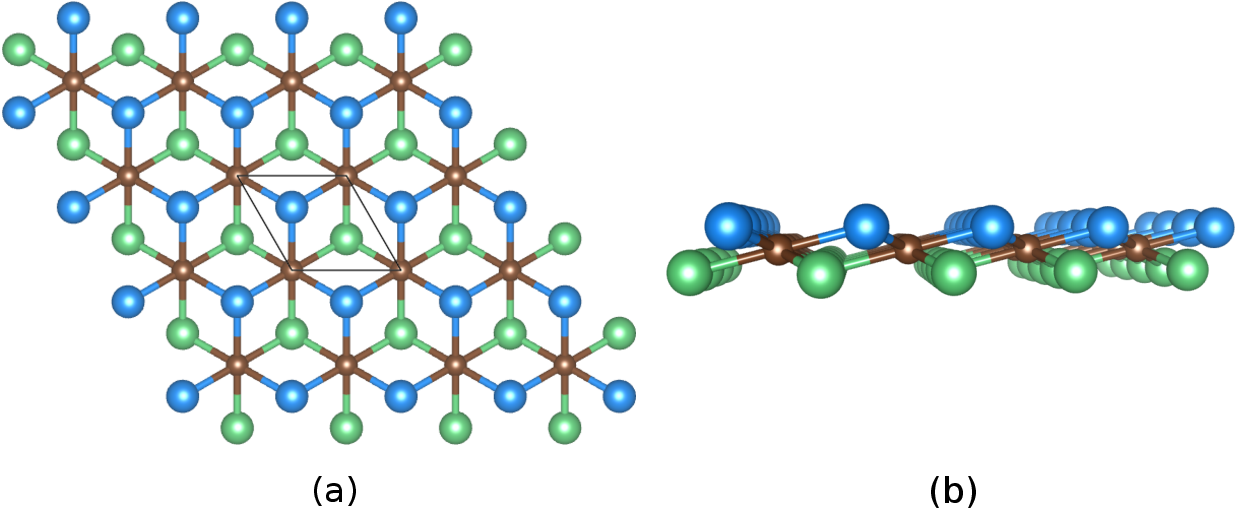}
\caption{(a) Top and (b) side view of the Be$_2$C-ML. Beryllium atoms are represented by dark blue (top) and light green (bottom) spheres and the carbon atoms in the middle layer are represented by brown spheres.}
\label{str}
\end{figure}

Be$_2$C exists in the hexagonal crystal structure with space group \textit{P}-3\textit{m1} (Fig. \ref{str}).
The Be$_2$C-ML is made of three atomic sublayers with C atoms sandwiched by two Be-layers stacked along the z-direction. Within Be$_2$C-ML, each C atom is connected with six Be atoms, and each Be atom is connected with three C, and three Be atoms. The optimized Be-C and Be-Be bond lengths are 1.76 {\AA} and 1.93 {\AA}, respectively. These values agree well with the previous report\cite{angwe_be2c}. The intralayer distance between the two Be layers is $\Delta$ $\sim$ 0.91 {\ AA} and that between the C and the Be layer is $\sim$ 0.45 {\ AA} as acquired from the present study.
Our calculations with PBE functional yield lattice parameter, a = 2.95 {\AA}, and it is in excellent agreement with the previously reported value 2.99{\AA}\cite{angwe_be2c}.
\subsection{Electronic band structure and density of states}
\begin{figure}[ht]
\centering
\includegraphics[scale=0.75]{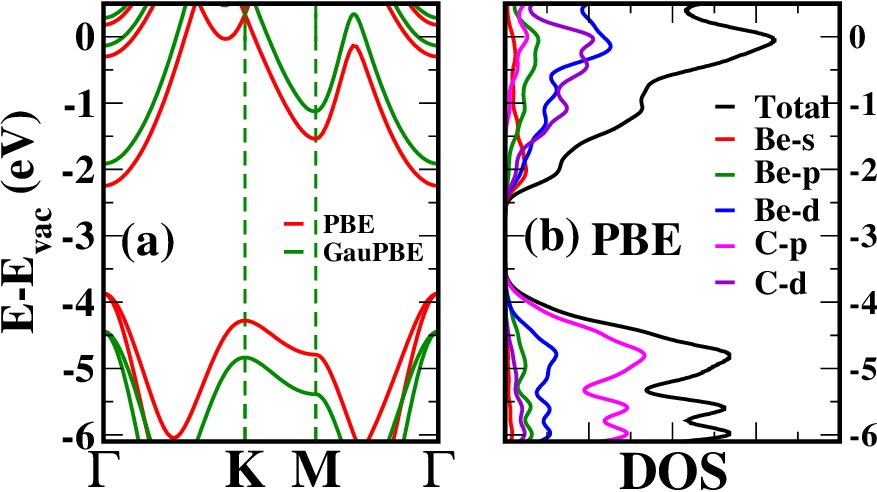}
\caption{(a) Band structure for Be$_2$C-ML with PBE (red) and Gau-PBE (green) functionals. (b) Projected density of states (PDOS) with PBE (b). The vacuum energy is chosen as zero to shift the Kohn-Sham energy eigenvalues.}
\label{bands_structure}
\end{figure}

The PBE band structure of Be$_2$C-ML (red solid lines) with vdW corrections is shown in Fig. \ref{bands_structure} (a). We observe a direct bandgap of 1.62 eV with the valence band maxima (VBM) and the conduction band minima (CBM) located at $\Gamma$ point of BZ. Since PBE functional underestimates the bandgap, we calculate the band structure with Gau-PBE hybrid functionals with vdW corrections to predict the accurate bandgap. These functionals provide Kohn-Sham eigenvalues as accurate as can be obtained by HSE hybrid functionals with very less computational cost\cite{gau}. We find that Gau-PBE predicts the bandgap of 2.0 eV, whereas the band dispersion remains intact compared to PBE as shown in Fig. \ref{bands_structure} (a) (solid green lines). The computed band gap is 0.34 eV lower than that previously reported with HSE\cite{angwe_be2c}.
The total density of states (DOS) and those projected onto the Be-$s,p,d$ and C-$p,d$ atomic orbitals are shown in Fig. \ref{bands_structure} (b). We see the valence band is primarily composed of the C-$p$ states, while the conduction band is
derived from the Be-$s,d$.

\subsection{Electron-Phonon relaxation time ($\tau$)}
The transport coefficients are produced in terms of relaxation time ($\tau$), namely, electrical ($\sigma$/$\tau$) and electronic thermal conductivity ($\kappa_{e}$/$\tau$) when computed within Boltzmann transport framework under CRTA. Therefore, it is of prime importance to calculate $\tau$ of the charge carriers explicitly to evaluate the absolute transport coefficients. Hence we exploit the most advanced method, EPW, based on the EPIs where acoustic and optical vibrational modes are taken to be scattering channels for the charge carriers. As a result, it precisely provides information of $\tau$ compared to traditionally used deformed potential theory wherein scattering only from acoustic modes is considered, and as a result, it severely overestimates the TE coefficients\cite{phosphorene_nano,phosphorene_prb, stanene,pccp_w}.
\begin{figure}[ht]
\centering
\includegraphics[scale=0.75]{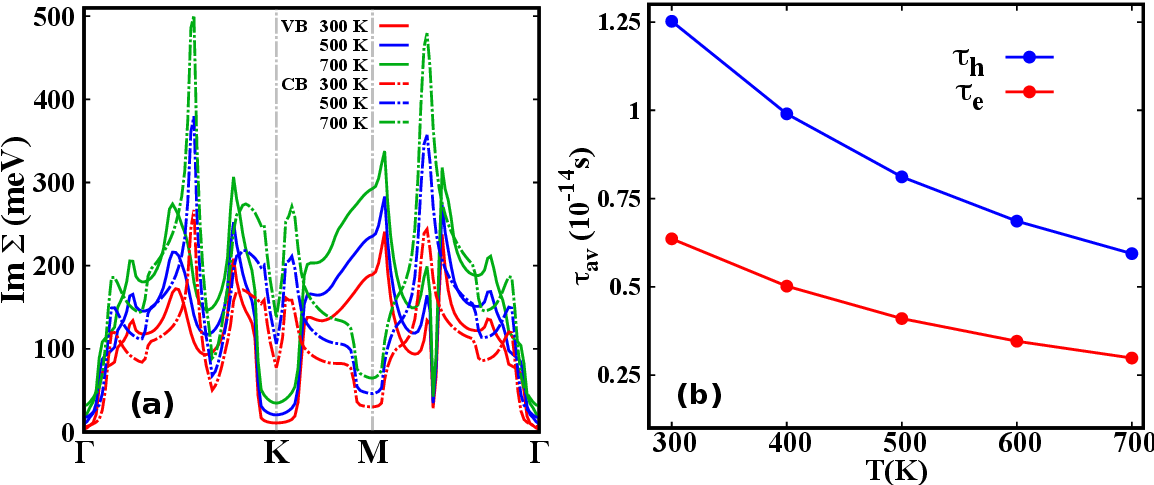}
\caption{(a) The variation of the imaginary part of the self-energy for the valence ($\Sigma_{VB}$) and the conduction bands ($\Sigma_{CB}$) (solid and dashed lines respectively) along the high symmetry directions in the BZ at 300K, 500K and 700K (red, blue and green respectively) for the Be$_2$C-ML. (b) Temperature-dependent average relaxation  time ($\tau_{av}$) for electron (red) and holes (blue).}
\label{epw_sigma}
\end{figure} 
Fig. \ref{epw_sigma} (a) shows the variation of the imaginary part of self-energy for the bottom-most conduction (solid lines) and the topmost valence bands (dashed lines) along the high symmetry directions in the BZ for the temperatures: 300K, 500K, and 700K at Fermi energy. We find that it increases with increased temperature due to increased scattering between excited phonons and electrons. As a result, it leads to a decrease in $\tau_{av}$ with temperature, as shown in
Fig. \ref{epw_sigma} (b). Here $\tau_{av}$ is obtained for a given band using ($\tau_{n}$ = $\sum\limits_{k}$ $\frac{\epsilon_{n,k} \tau_{n,k}}{\epsilon_{n,k}}$)\cite{tau_av}. Similarly, $\tau_{av}$ can also be obtained along the zigzag and armchair directions by supplying path along the M-K and $\Gamma$-M directions in BZ, respectively. Having the information of the $\tau_{av}$ for holes/electrons at various temperatures, we compute the transport properties of the Be$_2$C-ML as explained in the following section.
\subsection{Transport properties} 
Transport coefficients of Be$_2$C-ML are produced as a function of electron/hole doping at three different temperature values. Notice that we compute TE properties, namely, electrical conductivity ($\sigma$) and electronic thermal conductivity ($\kappa_{e}$) in the zigzag and armchair directions by incorporating $\tau_{av}$ computed in respective directions as mentioned in the previous section. Since the Seebeck coefficient ($\alpha$) is obtained independent of $\tau$ within
Boltzmann transport framework under CRTA. Therefore, it is expected to be the same in the two directions. Below, we discuss the transport properties along the zigzag direction and the same along the armchair direction discussed in Supplementary Material.

In Fig. \ref{TE_zz} (a), we plot the magnitude of the Seebeck coefficient ($|\alpha|$) as a function of carrier concentration at different temperatures for holes (solid lines) and electron doping (dashed lines). We find that the $\alpha$ decreases as the carrier concentration increases, and it increases with an increase in temperature, which is typical behavior of $\alpha$\cite{nature_snyder}.
The value of $\alpha$ (at optimum PF obtained with hole doping) is found to be 0.167 mV/K with hole doping of $\sim$ 6.6$\times$10$^{12}$ cm$^{-2}$ at 300 K and it is of the similar order as observed for doped-Bi$_2$Te$_3$\cite{pf_bi2te3_aip1,pf_bi2te3_aip2}. Notice that the $\alpha$ is greater for electrons than holes due to differences in effective masses of the carriers. The value of $\alpha$ for the electron doping is -0.167 mV/K at the carrier concentration of 1.2$\times$10$^{13}$ cm$^{-2}$ at 300 K where the PF tends to optimum value with electron doping. 

Fig. \ref{TE_zz} (b) depicts the electrical conductivity for the Be$_2$C-ML with hole (electron) doping. The $\sigma$ decreases with increased temperature since relaxation time decreases with temperature
for holes and electrons. We have obtained the value of $\sigma$ (at optimum PF) to be 1.22 $\times$ 10$^{5}$ S/m with hole doping of 6.6 $\times$ 10$^{12}$ cm$^{-2}$ at 300 K. Fig. \ref{TE_zz} (c) presents electronic contribution to thermal conductivity with hole/electron doping. We find that $\kappa_e$  is directly proportional to electrical conductivity, which is a typical trend as suggested by Wiedemann-Franz law. 

Fig. \ref{TE_zz} (d) shows the PF for the monolayer when it is hole/electron doped. The power factor is a function of $\alpha$ and $\sigma$ where both show opposite trends when the monolayer is doped. Therefore, one is expected to find the optimum value of PF as a function of carrier concentration. For hole doped of 6.6$\times$10$^{12}$ cm$^{-2}$, we have obtained the optimum value of PF to be $\sim$ 3.44 mW/mK$^2$ at 300 K which is of the same order observed for doped-Bi$_2$Te$_3$ and PbTe\cite{pf_bi2te3_aip1,pf_bi2te3_aip2,pbte_aip}.
Furthermore, it increases with temperature as $\alpha$ ramps up with temperature.
The value of optimum PF is 0.92 mW/mK$^2$ at the electron doping of 1.2$\times$10$^{13}$ cm$^{-2}$, and it is almost four times lower compared to PF obtained for hole doping due to the lower electrical conductivity of the electrons.

\begin{figure}[ht]
\centering
\includegraphics[scale=1]{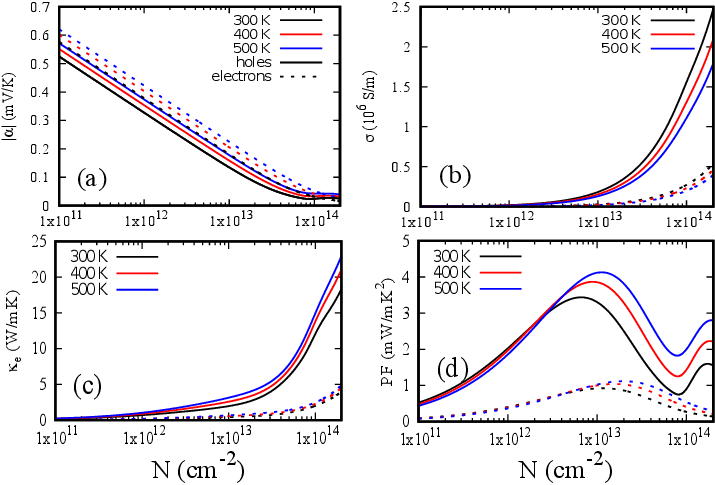}
\caption{The Seebeck coefficient (a) electrical conductivity (b) electronic thermal conductivity (c) power factor (d) with hole (electron) doping along the zigzag direction is shown by solid (dashed) lines  at 300 K, 400 K and 500K represented by black, red and blue lines respectively.}
\label{TE_zz}
\end{figure}

Interestingly, the high values of PF found in our case could challenge Bi$_2$Te$_3$ and PbTe\cite{pbte_aip,pf_bi2te3_aip1,pf_bi2te3_aip2}. A similar feature has also been observed in Heusler alloy like Fe$_2$VAl\cite{fe2val}. So our results can motivate further theoretical and experimental investigation on this material.

\section{CONCLUSION}
\label{concl}
In conclusion, we have computed the electronic and transport properties of Be$_2$C monolayer with Gau-PBE hybrid functional. Band structure calculations show that Be$_2$C-ML is a direct bandgap (E$_g$=2.0 eV) material with a doubly degenerate energy band at $\Gamma$ in the valence band. We have investigated electron-phonon interactions in the system to calculate the average relaxation time for charge carriers which is then incorporated in TE coefficients. The obtained value of PF ($\sim$ 3.44 mW/mK$^2$ at 300 K) is found to be of a similar order of PF as reported for many well-known TE materials like doped-Bi$_2$Te$_3$ and PbTe. Such a high value of PF reflects that Be$_2$C-ML may be a potential room temperature TE material.

\section{SUPPLEMENTAL MATERIAL}
Transport properties along armchair direction can be found in Supplementary Material.

\begin{acknowledgements}
GS would like to thank IISER-Pune for the fellowship. KCB acknowledges the DST-SERB for the SERB-National Postdoctoral Fellowship (Award No. PDF/2017/002876). The research used resources at the National Energy Research Scientific Computing Center (NERSC), which is supported by the Office of Science of the U.S. DOE under Contract No. DE-AC02- 05CH11231 and high performance computing Center of Development of Advanced Computing, Pune, India.
\end{acknowledgements}

\bibliography{ref}

\end{document}



\title{Supplementary Material: Impressive Transport Properties of Be$_2$C Monolayer} 



\author{Gautam Sharma}
\email{gautam.iiser@gmail.com}

\affiliation{Department of Physics, Indian Institute of  Science Education and Research Pune, Pune, India}
\author{K.C. Bhamu}
\affiliation{PMC Division, CSIR-National Chemical Laboratory, Pune-411008, India}

\maketitle
\section{Transport properties}
Transport properties are shown as a function of hole and electron doping along armchair direction for Be$_2$C monolayer.
\begin{figure}[ht]
\centering
\includegraphics[scale=1]{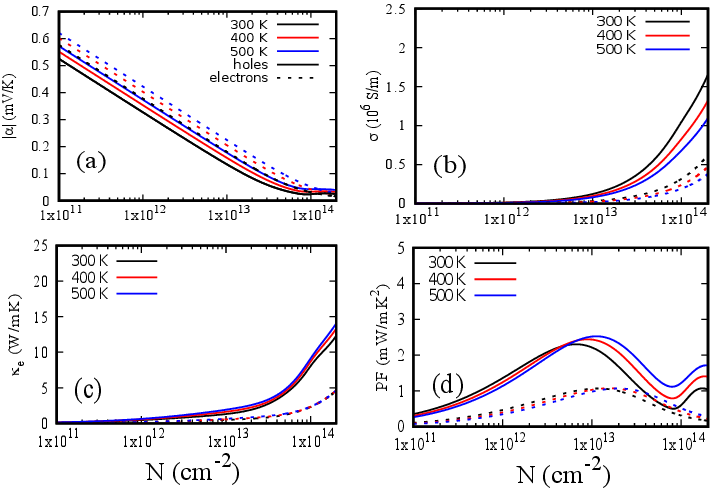}
\caption{The Seebeck coefficient (a) electrical conductivity (b) power factor (c) electronic thermal conductivity (d) with hole (electron) doping along armchair direction is shown by solid (dashed) lines  at 300 K, 400 K and 500K represented by black, red and blue lines respectively.}
\label{TE_arm}
\end{figure}